\documentclass[conference,compsoc]{IEEEtran}

\ifCLASSOPTIONcompsoc
  \usepackage[nocompress]{cite}
  \usepackage{graphicx}
  \usepackage{hyperref}
  \usepackage{footnote}
  \usepackage{algpseudocode}
  \usepackage{algorithm}
\else
  \usepackage{cite}
\fi

\usepackage{subfig}
\usepackage{amsmath}
\usepackage{multirow}
\usepackage{amsthm}
\newtheorem{theorem}{Theorem}
\newtheorem{corollary}[theorem]{Corollary}

\begin{document}

\title{Measuring DEX Efficiency and The Effect of an Enhanced Routing Method on Both DEX Efficiency and Stakeholders' Benefits}

\author{\IEEEauthorblockN{Yu Zhang} 
\IEEEauthorblockA{BDLT, IfI Department, \\University of Zurich\\ Zurich, Switzerland \\Email: zhangyu@ifi.uzh.ch}
\and
\IEEEauthorblockN{Claudio J. Tessone} 
\IEEEauthorblockA{BDLT, IfI Department, \\University of Zurich\\ Zurich, Switzerland}
}
\maketitle

\begin{abstract}
The market efficiency of decentralized exchanges (DEX), as well as the impacts of different token routing algorithms on DEX efficiency and stakeholders' benefits, has not been thoroughly explored by researchers. In this paper, we first put forward the concept of standardized total arbitrage profit (STAP) calculated by the convex optimization method to measure the DEX efficiency systematically. Through mathematical proof, it is demonstrated that no arbitrage opportunities remain—even after executing the trading order derived from maximizing the STAP and reintegrating the retained transaction fees into their respective liquidity pools. This conclusion holds for both cyclic arbitrage opportunities within decentralized exchanges (DEXs) and arbitrage opportunities between DEXs and centralized exchanges (CEXs). In an efficient decentralized exchange (DEX) where the STAP is zero, the monetized value of the received target tokens must be less than or, at most, equal to the monetized value of the source tokens, regardless of the routing algorithms employed by traders. Any deviation from this condition implies the presence of arbitrage opportunities. This criterion may therefore serve as a standard for evaluating the existence of arbitrage opportunities within a DEX. By measuring STAP on a token graph with 11 tokens and 18 liquidity pools with data from Uniswap V2, we find that the efficiency of Uniswap V2 decreases from June 21, 2024, to November 8, 2024. By simulating token trading using two routing algorithms—the line-graph-based method proposed by \cite{zhang2025line} and the depth-first search (DFS) algorithm—it is observed that employing a more profitable routing strategy can enhance the overall efficiency of the DEX and improve traders' benefits as a whole with time. Furthermore, the total value locked (TVL) within the DEX remains relatively stable throughout the simulations if the line-graph-based method is used in token routing. While the TVL increases if the DFS algorithm is used in routing, which means an increase in all liquidity providers' benefits as a whole.
\end{abstract}

\IEEEpeerreviewmaketitle

\section{Introduction} 

In 2008, the global financial crisis induced the distrust of centralized financial institutions. In 2009, the Bitcoin blockchain platform was launched based on the paper entitled `Bitcoin: A Peer-to-Peer Electronic Cash System' \cite{nakamoto2008peer} by Satoshi Nakamoto and attracted much interest from individuals, especially computer scientists, developers who initially encountered this technology \cite{sornette2025transaction}. Now, Bitcoin has been one of the most successful decentralized blockchain platforms in payment, and more and more governments, academic researchers, and financial industry specialists are exploring blockchain's new applications in finance. In 2014, another significant and successful blockchain platform, Ethereum, was established with a distinct characteristic, namely, it supports smart contracts, which are a code snippet that can be defined by users and run on the blockchain. Due to the introduction of smart contracts by Ethereum, we are stepping into the decentralized finance (DeFi) era, where the decentralized exchange (DEX) plays a critical role in token swapping and trading.

A DEX consists of numerous liquidity pools, each containing various types and quantities of tokens, which are supplied by liquidity providers. Any individual may become a liquidity provider by either creating a new liquidity pool or contributing tokens to an existing one. Token trading on decentralized exchanges occurs within liquidity pools governed by automated market-making (AMM) protocols. These protocols include models such as the Constant Product Market Maker (CPMM), Constant Sum Market Maker (CSMM), Constant Mean Market Maker (CMMM), among others \cite{gogol2024sok}. The CPMM model is the most widely adopted, employing a constant product formula to determine exchange prices. This formula maintains that the product of the quantities of two assets within a liquidity pool remains constant. It has been extensively implemented by major decentralized exchanges, including Uniswap, PancakeSwap, SushiSwap, and Balancer.

Much research has focused on the arbitrage detection in DEXs, the statistics of arbitrage transactions in DEXs, and the token routing algorithm design for token trading in DEXs. However, there is still little research about how to measure the efficiency of DEXs systematically and how significant factors, like the token routing methods, affect the DEX efficiency and related stakeholders' benefits. 
In this paper, we will try to explore these research questions based on Uniswap V2's CPMM model and data to provide a bit of insight about how the DEX can be improved in the future. 

The paper is organized as follows:
In Section \ref{related_work}, related research about DEX efficiency is introduced. In Section \ref{efficiency_measurement}, we put forward the concept of standardized TAP (total arbitrage profit in the token graph) to measure the efficiency of DEX systematically and prove several significant characteristics in mathematics. In Section \ref{data_description}, we briefly introduced the data we used for measuring the DEX efficiency. In Section \ref{efficiency_measure_token_routing}, we measure the efficiency of Uniswap V2 based on the data from June 21, 2024, to November 8, 2024, and simulate how different token routing algorithms affect the efficiency of DEX and related stakeholders' benefits. At last, we summarize the paper and discuss the limitations of the research and potential work in the future. 

\section{Related Work}\label{related_work}

In DEX-related research, much research focuses on arbitrage statistics using historic transaction data or arbitrage opportunities detection with different methods. For example, \cite{wang2022cyclic} conduct an analysis of potential cyclic arbitrage opportunities by systematically traversing all triangular trading paths that involve Ether, aiming to quantify the associated arbitrage profits. \cite{mclaughlin2023large} extract arbitrage transactions from historical trade event logs and utilize Johnson’s cycle-detection algorithm to systematically identify potential arbitrage opportunities.
\cite{danos2021global} took the arbitrage problem as a convexity problem and applied the convex optimization operation to detect arbitrage opportunities from a theoretical perspective.
\cite{zhang2024improved} designed a line-graph-based method to detect more arbitrage opportunities in DEX, like Uniswap V2. Arbitrage opportunities can reflect the efficiency of DEX, but are not a perfect indicator to measure DEX efficiency systematically because it is hard to enumerate all arbitrage opportunities in a DEX.

There is also research that focuses on token routing in DEX. For example, \cite{danos2021global} converts the token routing problem to a convex optimization problem. While the drawback of using the convex optimization method is that its computational complexity is extremely high, and the calculated result is almost impossible to execute, which is shown in \cite{zhang2025line}. Then, \cite{zhang2025line} designed a line-graph-based method to detect a routing path for token trading and found that it is more profitable than the popularly used DFS algorithm in the linear case.

Although the market efficiency has been researched a lot in the traditional financial market, there is limited research on the efficiency of DEX and how the token routing algorithms affect it, and the related stakeholders' benefits.
In DEX efficiency, we only find that \cite{berg2022empirical} discusses the market inefficiency of decentralized exchanges (DEXs) and examines their inefficiencies within Uniswap V2 by identifying more profitable token trading paths through route-splitting techniques and detecting profitable cyclic arbitrage opportunities. How to systematically measure the DEX efficiency and research how the token routing algorithm affects it and the related stakeholders' benefits still needs more exploration.

\section{Measuring The Efficiency of DEX Systematically}\label{efficiency_measurement}

The paper focuses on the Uniswap V2 DEX because it is simpler to model, was the largest DEX, and is still significant and active in trading. The research result can be extended to other DEXs that use different and more complicated AMM rules.

\begin{figure}[htbp]
\subfloat[Trading curve of CPMM on Uniswap V2]{\centering
    \includegraphics[width=0.5\linewidth]{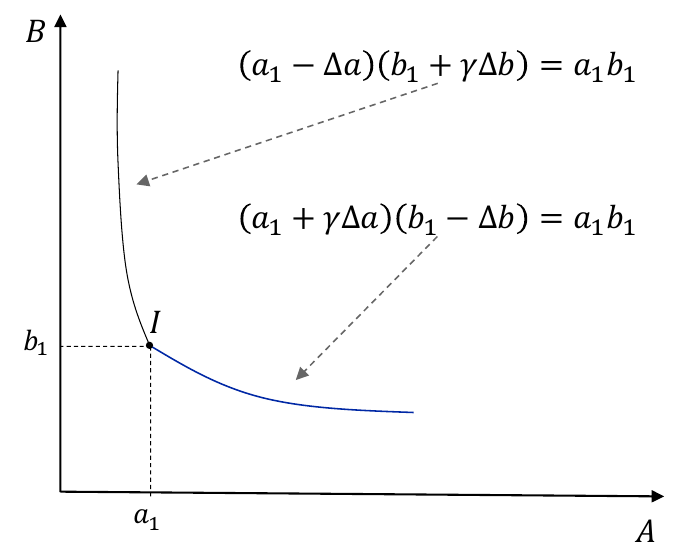}
    \label{cpmm_intro}}
\subfloat[Trading phases on Uniswap V2]{ \centering
    \includegraphics[width=0.5\linewidth]{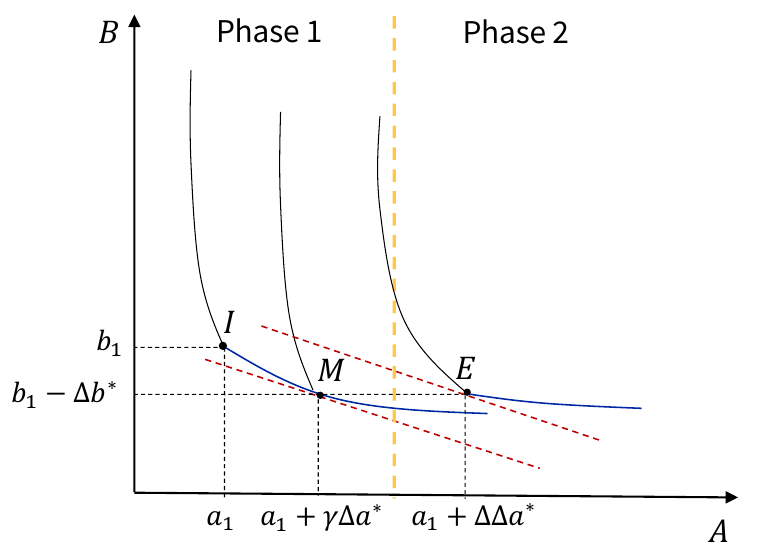}
    \label{cpmm_intro1}}
\caption{Token trading curve and two phases during token trading in liquidity pools on Uniswap V2. In Fig.\ref{cpmm_intro}, $a_1$ and $b_1$ are the initial token reserves of token $A$ and $B$. The blue line to the right of point $I$ represents selling token $A$ to purchase token $B$, while the black line to the left of point $I$ represents selling token $B$ to purchase token $A$.  The red line in Fig.\ref{cpmm_intro1} denotes the ratio of two tokens' CEX prices, namely $\frac{P_A}{P_B}$. $P_A$, $P_B$ are token $A$ and $B$'s price from some CEX.}
\label{cpmm_curve}
\end{figure}

Uniswap V2 was launched in May 2020 and uses the constant product market making (CPMM) to manage the token trading activities automatically that take place in the token liquidity pools, namely $a\cdot b =k$ where $k$ is the constant product of tokens' reserve in a liquidity pool, $a$ and $b$ are the corresponding tokens' (token $A$ and token $B$) reserves in the same liquidity pool. Assuming tokens' initial reserves are $a_1$ and $b_1$, the token trading formula are:
\begin{equation}
    \begin{aligned}
        &(a_1+\gamma \Delta a)\cdot(b_1 - \Delta b) = a_1\cdot b_1 \quad selling \ A\\
        &(a_1-\Delta a)\cdot(b_1 +\gamma \Delta b) = a_1\cdot b_1 \quad selling\ B\\
    \end{aligned}
    \label{cpmm}
\end{equation}
where $\gamma = 1-\lambda$ and $\lambda$ (0.3\% in Uniswap V2) is the transaction fee rate. Its curve is shown in Fig.\ref{cpmm_intro}

As we can see from Fig.\ref{cpmm_intro}, the initial reserves of token $A$ and $B$ are $a_1$ and $b_1$, respectively, namely the point of $I$. The horizontal axis denotes token $A$'s reserve, and the vertical axis denotes token $B$'s reserve. The line in the left of point $I$ denotes the case that traders buy token $A$ with token $B$, and the line in the right side of point $I$ denotes the case that users buy token $B$ by selling token $A$ to the liquidity pool.

After calculating the number of target tokens based on the CPMM rule as shown above in equation \ref{cpmm} (that are $\Delta b$ in the first equation and $\Delta a$ in the second equation), the retained transaction fee (that are $\lambda \Delta a$ in the first equation and $\lambda \Delta b$ in the second equation) will go to the liquidity pool again as a reward to liquidity providers.
So, the whole procedure of token trading in a liquidity pool should include two phases. In the first phase, the transaction fee is retained, the number of tokens to be received is calculated according to the Constant Product Market Maker (CPMM) formula, and the token exchange is then executed. In the second phase, the retained fee is added to the liquidity pool.

Following the above analysis, assuming we buy token $B$ by selling token $A$, the two phases are shown in Fig.\ref{cpmm_intro1}. In the first phase, we calculate the number of token $B$ we can receive by inputting a specific number of token $A$. Then we execute the trading, which is reflected by the curve movement from point $I$ to point $M$. In the second phase, we add the retained transaction fee ($\lambda \cdot \Delta a$) to the liquidity pool, which is reflected by the curve movement from point $M$ to $E$.

A DEX usually contains plenty of liquidity pools and owns a very complex token link structure, resulting in tokens' price disparity. 
Arbitrageurs pursue the maximal arbitrage profit if they find an arbitrage opportunity. Fewer arbitrage opportunities usually mean that the market is efficient. A natural idea to measure the efficiency of DEX is to calculate its total arbitrage profit existing in the considered DEXs.
In DEX, the functions of Auto-market makers (AMM), such as the constant product market making (CPMM) rule in Uniswap V2, are convex. Then, the total arbitrage profit maximization problem can be converted to a convex optimization problem, as shown in equation \ref{optimization}.

\begin{equation}
\begin{aligned}
\max \quad & P_A\sum_i(\Delta a_{out}^i - \Delta a_{in}^i)
+P_B\sum_i(\Delta b_{out}^i - \Delta b_{in}^i)+\cdots\\
\textrm{s.t.} \quad 
&(a_1+\gamma \Delta a_{in}^1-\Delta a_{out}^1)\cdot(b_1 +\gamma \Delta b_{in}^1- \Delta b_{out}^1) \geq a_1\cdot b_1\\
&(a_2+\gamma \Delta a_{in}^2-\Delta a_{out}^2)\cdot(c_2 +\gamma \Delta c_{in}^2- \Delta c_{out}^2) \geq a_2\cdot c_2
&\\
\cdots\\
& \Delta a_{out}^i, \Delta a_{in}^i, \Delta b_{out}^i, \Delta b_{in}^i,\cdots, \geq 0.
\end{aligned}
\label{optimization}
\end{equation}
where $\gamma=1-\lambda$ and $\lambda$ is transaction fee rate. $a_i$, $b_i$, $c_i$, $\cdots$ are the reserves of token $A$, $B$, $C$, $\cdots$ in the $i^{th}$ liquidity pool. For example, in the first pool, token $A$ and $B$'s reserves are $a_1$ and $b_1$. All other tokens' reserves are zero in the first liquidity pool because there are only tokens $A$ and $B$ in this pool. $\Delta T_{in}^i$ and $\Delta T_{out}^i$ denote token $T$'s input and output to the $i^{th}$ liquidity pool, with $T=A, B, C,\cdots$. $P_T\cdot \sum_i(\Delta T_{out}^i - \Delta T_{in}^i)$ measures how much profit we can obtain by selling and buying token $T$. $P_T$ is the token $T$'s price from some centralized exchange. So, the total summation $TAP= P_A\sum_i(\Delta a_{out}^i - \Delta a_{in}^i)
+P_B\sum_i(\Delta b_{out}^i - \Delta b_{in}^i)+\cdots$ measures the total arbitrage profit that exists in the corresponding token graph, with $TAP$ being an abbreviation of the phrase `total arbitrage profit'.

When $TAP$ is high, then the corresponding DEX may not be efficient because many arbitrage opportunities exist. Assuming that each token's price increased by the same percent $q$ with all tokens' reserve in corresponding liquidity pools unchanged, we suppose to expect that the efficiency of the corresponding DEX does not change. However, $TAP$ will also increase by the same percentage $q$ when all tokens' prices increase by $q$, which means that there are drawbacks in using $TAP$ to measure the efficiency of DEX. This problem can be solved by dividing the TVL of all liquidity pools in the corresponding token graph by $TAP$, namely, $\frac{TAP}{TVL}$, with $TVL= P_A\cdot\sum_i a_i+P_B\cdot\sum_i b_i+\cdots$. 
In summary, the indicator to measure the efficiency of DEX can be:
$STAP = \frac{P_A\sum_i(\Delta a_{out}^i - \Delta a_{in}^i)
+P_B\sum_i(\Delta b_{out}^i - \Delta b_{in}^i)+\cdots}{P_A\cdot\sum_i a_i+P_B\cdot\sum_i b_i+\cdots}$, which can be called standardized $TAP$.

For the optimization problem \ref{optimization}, it can be proved that there are no cyclic arbitrage opportunities in the DEX, and no arbitrage opportunities between DEX and CEX in the optimal point where the objective function is maximal.
\begin{theorem}
    Assuming $r^*$ is the point where the objective function $\frac{TAP}{TVL}$ is maximal. Because this is a convex optimization problem, we know that the optimal point $r^*$ is unique. In the optimal point of the convex optimization problem \ref{optimization}, there are no cyclic arbitrage opportunities anymore in the corresponding DEX. 
\label{theorem1}
\end{theorem}

\begin{proof}
    Assuming there is a cyclic arbitrage opportunity, which is $A\rightarrow B\rightarrow C\rightarrow A$. $A$, $B$ and $C$ are tokens. In this cyclic arbitrage opportunity, for any token from $A$, $B$ and $C$, we can always calculate the optimal input count $inp$ and corresponding output count $oup$ so that $oup-inp$ is maximal and $oup-inp>0$. During the trading of tokens along the cyclic arbitrage loop, all constraints are still satisfied. However, the objective function will increase by $P\cdot(oup-inp)$, where $P$ is the corresponding token's price. This result contradicts to the premise that the objective function is already maximal. So, cyclic arbitrage opportunities do not exist anymore in the optimal point of the convex optimization problem.
\end{proof}

\begin{theorem}
    In the optimal point $r^*$, there are also no arbitrage opportunities between DEX and centralized exchanges (CEXs) for any pair of tokens.
    \label{theorem2}
\end{theorem}

\begin{proof}
    We convert the convex optimization problem to a Lagrange problem, which is:
    \begin{equation}
    \begin{aligned}
&F=P_A\sum_i(\Delta a_{out}^i - \Delta a_{in}^i)
+P_B\sum_i(\Delta b_{out}^i - \Delta b_{in}^i)+\cdots\\
&+L_1\{(a_1+\gamma \Delta a_{in}^1-\Delta a_{out}^1)\cdot(b_1 +\gamma \Delta b_{in}^1- \Delta b_{out}^1)- a_1\cdot b_1\}\\
&+L_2\{(a_2+\gamma \Delta a_{in}^2-\Delta a_{out}^2)\cdot(c_2 +\gamma \Delta c_{in}^2- \Delta c_{out}^2) -a_2\cdot c_2\}\\
&+\cdots.
    \end{aligned}
    \end{equation}
We focus on the first constraint which corresponds to the liquidity pool containing $a_1$ units of token $A$ and $b_1$ units of token $B$. 
Then we calculate the derivative of $F$ to all variables, including $\Delta a_{out}^i$, $\Delta b_{in}^i$, etc..
\begin{equation}
\begin{aligned}
&\frac{\partial F}{\partial \Delta a_{in}^1} =- P_A + L_1\gamma(b_1 +\gamma \Delta b_{in}^1- \Delta b_{out}^1)=0\\
&\frac{\partial F}{\partial \Delta b_{out}^1} = P_B - L_1(a_1+\gamma \Delta a_{in}^1-\Delta a_{out}^1)=0\\
&\cdots
\end{aligned}
\label{derivative_eq}
\end{equation}
Because this is a convex optimization problem, the optimal solution must be on the boundary of the constraints. 
Namely $(a_1+\gamma \Delta a_{in}^1-\Delta a_{out}^1)\cdot(b_1 +\gamma \Delta b_{in}^1- \Delta b_{out}^1)- a_1\cdot b_1=0$ and either $\Delta a_{in}^1=\Delta b_{out}^1=0$ or $\Delta a_{out}^1=\Delta b_{in}^1=0$.

We assume that $\Delta a_{out}^1=\Delta b_{in}^1=0$, then $\Delta a_{in}^1 \geq0$ and $\Delta b_{out}^1\geq 0$, which means we sell token $A$ to buy token $B$. Then, the above equation \ref{derivative_eq} will be simplified as the following equation \ref{derivative_eq_simplified}:
\begin{equation}
\begin{aligned}
&\frac{\partial F}{\partial \Delta a_{in}^1} =- P_A + L_1\gamma(b_1 - \Delta b_{out}^1)=0\\
&\frac{\partial F}{\partial \Delta b_{out}^1} = P_B - L_1(a_1+\gamma \Delta a_{in}^1)=0\\
&\cdots
\end{aligned}
\label{derivative_eq_simplified}
\end{equation}
Then we have $\frac{P_A}{P_B}=\frac{\gamma(b_1 - \Delta b_{out}^{1*})}{a_1+\gamma\Delta a_{in}^{1*}}$. By solving this convex optimization problem, we can calculate the number of each token input and output, namely these variables $\Delta \cdot$, which corresponds to the curve movement from point $I$ to $M$ in the first phase of token trading in the Uniswap V2 liquidity pools, as shown in Fig.\ref{cpmm_intro1}. 

We can find that tokens' relative prices in the DEX are equal to those in CEXs in $M$ as shown in Fig.\ref{cpmm_intro1}, which means that no arbitrage opportunities exist between DEX and CEX by selling token $A$ to buy token $B$ in the optimal point. On the other side of the optimal point where we can only buy token $A$ by selling token $B$, there is also no arbitrage profit. In summary, there are no arbitrage opportunities between DEX and CEX at the optimal point where the objective function is maximal.
\end{proof}

Theorems \ref{theorem1} and \ref{theorem2} apply to the first phase during token trading in the Uniswap V2 liquidity pools. After the first phase, we still need to add the retained transaction fees to the liquidity pools as a reward to liquidity providers, which is the second phase of token trading in Uniswap V2's liquidity pool. Adding these retained transaction fees to the corresponding liquidity pools will change the reserve ratio, which lead to $\frac{P_A}{P_B}\neq\frac{\gamma(b_1 - \Delta b_{out}^{1*})}{a_1+\gamma\Delta a_{in}^{1*}}$. Then, will new arbitrage opportunities show up, including the cyclic arbitrage opportunities in DEX, and the arbitrage opportunities between DEX and CEX?
In the following part, we will prove that no new arbitrage opportunities show up even after the retained transaction fees are added to the corresponding liquidity pools in the second phase of token trading in Uniswap V2.

\begin{theorem}
    After the second phase of token trading in Uniswap V2's liquidity pool, there are no arbitrage opportunities by buying and selling tokens between DEX and CEX, even if the token reserve ratios in any DEX's liquidity pools don't equal the ratios of their corresponding prices from CEX.
    \label{theorem3}
\end{theorem}
\begin{proof}
    Assuming we need to buy $\Delta b*$ token $B$ by selling $\Delta a*$ units of token $A$ in some liquidity pool containing both token $A$ and $B$, which corresponds to point $M$ in Fig.\ref{cpmm_intro1}. 
    At point $M$, the trading functions on both sides are:
    \begin{equation}
    \begin{aligned}
        &(a_1+\gamma \Delta a^*+ \gamma \Delta a)\cdot(b_1 - \Delta b^*-\Delta b) = a_1\cdot b_1 \quad selling \ A\\
        &(a_1+\gamma \Delta a^*-\Delta a)\cdot(b_1 - \Delta b^* +\gamma \Delta b) = a_1\cdot b_1 \quad selling\ B\\
    \end{aligned}
    \label{cpmm_proof3}
\end{equation}
where $\Delta a$ and $\Delta b$ are variables, all others are constant numbers.
Based on the analysis above, we know that the slope calculated from the right side of point $M$ equals $\frac{P_A}{P_B}$ and 
$\frac{d\Delta a}{d\Delta b}\mid_M^+ = \frac{\gamma(b_1 - \Delta b^*-\Delta b)}{a_1+\gamma \Delta a^*+ \gamma \Delta a}$
At $M$, we have $\Delta a=\Delta b =0$, then $\frac{P_A}{P_B}=\frac{d\Delta a}{d\Delta b}\mid_M^+= \frac{\gamma(b_1 - \Delta b^*)}{a_1+\gamma \Delta a^*}$

The slope calculated from the left side of point $M$ equals
$\frac{d\Delta a}{d\Delta b}\mid_M^- = \frac{b_1 - \Delta b^* +\gamma \Delta b}{\gamma(a_1+\gamma \Delta a^*-\Delta a)}$. At $M$, we have $\Delta a=\Delta b =0$, then, $\frac{d\Delta a}{d\Delta b}\mid_M^- = \frac{b_1 - \Delta b^*}{\gamma(a_1+\gamma \Delta a^*)}$

We can find that $\frac{d\Delta a}{d\Delta b}^->\frac{d\Delta a}{d\Delta b}\mid_M^+=\frac{P_A}{P_B}$, namely, traders can not arbitrage by selling token $B$ to buy token $A$, as proved in theorem \ref{theorem2}.

Now, if we add the retained transaction fee $(1-\gamma)\Delta a^*$ to the liquidity pool that containing both token $A$ and $B$, then the new trading functions are:
\begin{equation}
    \begin{aligned}
        &(a_1+\Delta a^*+ \gamma \Delta a)\cdot(b_1 - \Delta b^*-\Delta b) = (a_1+\lambda \Delta a^*)\cdot b_1 \quad selling \ A\\
        &(a_1+\Delta a^*-\Delta a)\cdot(b_1 - \Delta b^* +\gamma \Delta b) = (a_1+\lambda \Delta a^*)\cdot b_1 \quad selling\ B\\
    \end{aligned}
    \label{cpmm_proof3-1}
\end{equation}
where $\lambda=1-\gamma$ and $\lambda$ is the transaction fee rate charged by liquidity pools. These two new functions correspond to the two curves on both sides of point $E$, as shown in Fig.\ref{cpmm_intro1}.

Now, we calculate their slopes at point $E$ from both sides of it.
Firstly, from the right side of $E$, we have $\frac{d\Delta a}{d\Delta b}\mid_E^+ = \frac{\gamma(b_1 - \Delta b^*-\Delta b)}{a_1+ \Delta a^*+ \gamma \Delta a}$
At $E$, we have $\Delta a=\Delta b =0$, then $\frac{d\Delta a}{d\Delta b}\mid_E^+=\frac{\gamma(b_1 - \Delta b^*)}{a_1+ \Delta a^*}$

Secondly, from the left side of $E$, we have:$\frac{d\Delta a}{d\Delta b}\mid_E^- = \frac{b_1 - \Delta b^* +\gamma \Delta b}{\gamma(a_1+ \Delta a^*-\Delta a)}$.
At $E$, we have $\Delta a=\Delta b =0$, then $\frac{d\Delta a}{d\Delta b}\mid_E^-=\frac{b_1 - \Delta b^*}{\gamma(a_1+ \Delta a^*)}$.

Based on the above analysis, we have:
\begin{equation}
    \begin{aligned}
       & \frac{d\Delta a}{d\Delta b}\mid_E^+=\frac{\gamma(b_1 - \Delta b^*)}{a_1+ \Delta a^*}< \frac{P_A}{P_B}=\frac{\gamma(b_1 - \Delta b^*)}{a_1+\gamma \Delta a^*}\\
       & \frac{d\Delta a}{d\Delta b}\mid_E^-=\frac{b_1 - \Delta b^*}{\gamma(a_1+ \Delta a^*)}> \frac{P_A}{P_B}=\frac{\gamma(b_1 - \Delta b^*)}{a_1+\gamma \Delta a^*}
    \end{aligned}
\end{equation}
So, traders also can not arbitrage by selling token $B$ to buy token $A$.
The result applies to all liquidity pools. So, there are no arbitrage opportunities between DEX and CEX for any liquidity pool.
\end{proof}

In theorem \ref{theorem1}, we proved that there are no cyclic arbitrage opportunities after executing the trading order calculated by maximizing the total arbitrage profit (TAP) in DEX, which is the first phase as shown in Fig.\ref{cpmm_intro1}. Now we prove that there are also no cyclic arbitrage opportunities in the DEX after the second phase when the retained transaction fees are added to the liquidity pool again. 

\begin{theorem}
    After the second phase of token trading in Uniswap V2's liquidity pools, there are still no cyclic arbitrage opportunities in the DEX.
    \label{theorem4}
\end{theorem}

\begin{proof}
    Assuming any token loop, such as a token loop with three tokens $A\rightarrow B\rightarrow C\rightarrow A$ for simplicity. Based on Theorem \ref{theorem3}, we know that: for the liquidity pool containing token $A$ and $B$, traders can not arbitrage between buying tokens in this pool and then selling to a CEX. This indicates that: if we input $\Delta a_{in}$ units of token $A$ and obtain $\Delta b_1$ units of token $B$ in the liquidity pool using the CPMM rule, then there must be the following formula $\Delta a_{in} \cdot P_A>\Delta b_1 \cdot P_B$. Similarly, for other pools along the trading loop, we have the following formulas:
    \begin{equation}
    \begin{aligned}
       & \Delta a_{in} \cdot P_A>\Delta b_1 \cdot P_B \\
       & \Delta b_1 \cdot P_B>\Delta c_1 \cdot P_C\\
       & \Delta c_1 \cdot P_C>\Delta a_{out} \cdot P_A
    \end{aligned}
\end{equation}
$a_{in}$ is the number of token $A$ we input to the trading loop, and $a_{out}$ is the number of token $A$ we received after trading along the loop.
We find that $\Delta a_{in} \cdot P_A>\Delta a_{out} \cdot P_A$, namely, $\Delta a_{in} >\Delta a_{out}$. This means the cyclic arbitrage opportunities are not possible.
The above analysis applies to any trading loops, which means that there are no cyclic arbitrage opportunities in DEX even after the retained transaction fees are added to the liquidity pool as a reward to the liquidity pool providers.
\end{proof}

These theorems and proofs also indicate that the TAP and standardized TAP may be a good measurement of the DEX efficiency.

\begin{figure}[htbp]
    \centering
    \includegraphics[width=0.7\linewidth]{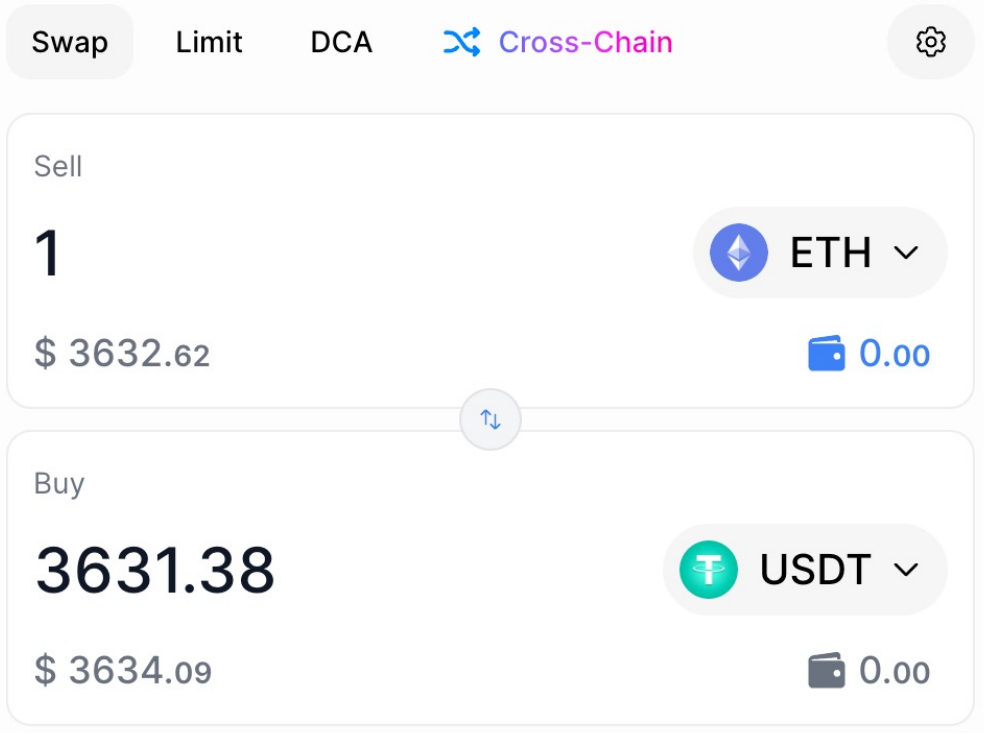}
    \caption{The trading interface of Sushiswap. At around 11:00 am, on July 24, 2024, the monetized value of 1 ETH to sell is 3632.62\$, while the monetized value of the target token USDT to buy is 3634.09\$. }
    \label{sushiswap_dex_interface}
\end{figure}

Based on the above four theorems, we have the following corollary:
\begin{corollary}
    In an efficient DEX where no cyclic arbitrage opportunities and no arbitrage opportunities between DEX and CEX, the monetized value of the received target token is less than or at most equal to the monetized value of the source token in token trading, whatever routing methods traders use.
\end{corollary}
In a DEX, if the monetized output value of the received target token is larger than the monetized value of the source token input, there must be arbitrage opportunities, either cyclic arbitrage opportunities in DEX or arbitrage opportunities between DEX and CEX. For example, we tried to trade 1 ETH for USDT on the Sushiswap platform, and found that the monetized value of the source token ETH is less than the monetized value of the received target token USDT as shown in Fig.\ref{sushiswap_dex_interface}, which means that the Sushiswap platform's efficiency is not high enough and arbitrage opportunities exist.

\section{Data Description}\label{data_description}

The pool reserves' data is from Uniswap V2 on the Ethereum blockchain, ranging from June 21, 2024, to November 8, 2024. 
Cryptocurrencies' daily price time series data is from CoinGecko. In CoinGecko, we can only download tokens' price data dating back to around one year ago without a subscription. Our downloaded tokens' price time series range from June 21, 2024, to June 20, 2025. In CoinGecko, each token has a unique token ID. The same token traded in different platforms has the same token ID. For example, wrapped bitcoins (WBTC) can be traded on both Ethereum and the BNB chain, but have different token smart contract addresses in the two blockchains. However, in CoinGecko, the two kinds of WBTC have the same token ID.
By token ID, users can also can also download the price time series of the corresponding tokens. Firstly, we map the tokens' smart contract addresses to their token IDs. Then, we use token IDs to download the tokens' price time series. By this way, we map tokens' smart contract addresses to their price time series from CEX.

During the construction of the token graph with Uniswap V2 pool reserve data, we first select pools whose total value locked (TVL) is larger than ten thousand dollars. Each pool data contains several items, such as pool address, two tokens' smart contract addresses, two tokens' reserves, and the timestamp when the pool reserve information is updated. Secondly, we select pools where both tokens' price time series data can be downloaded from CoinGecko. 

\begin{figure}[htbp]
    \centering
    \includegraphics[width=1\linewidth]{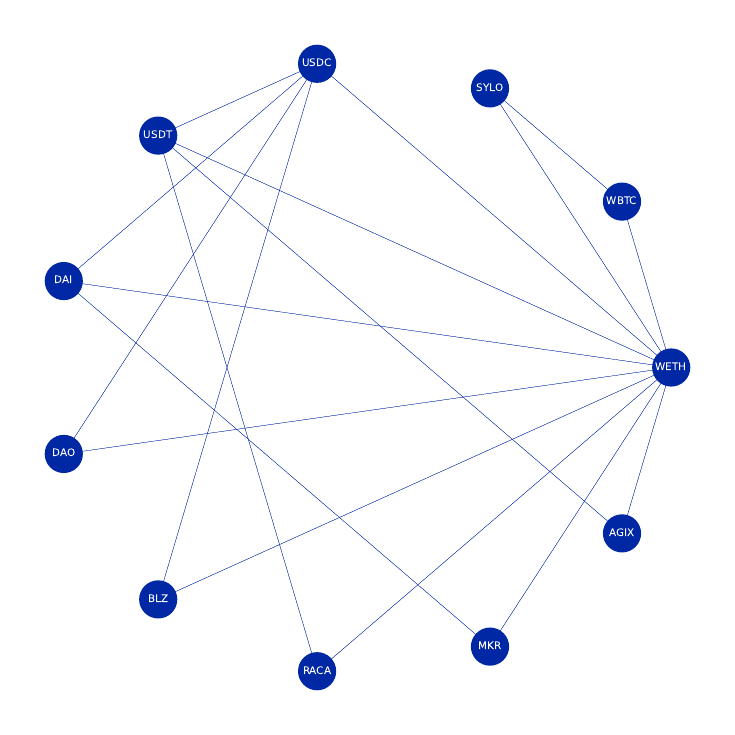}
    \caption{Token graph. Each node denotes a token, and each link denotes a liquidity pool containing both tokens at the ends of the corresponding link. There are 11 tokens (nodes) and 18 liquidity pools (edges) in this token graph. The 11 tokens are: SYLO, WBTC, WETH, AGIX, MKR, RACA, BLZ, DAO, DAI, USDT, and USDC.}
    \label{token_graph}
\end{figure}

After the token graph is constructed with nodes denoting tokens that are identified by their smart contract addresses, and edges denoting liquidity pools containing both tokens at the ends of the edges, we remove nodes whose degree is less than two iteratively. The step is similar to \cite{zhang2024improved}. At last, we remove pools with the lowest TVL one by one, until there are only around twenty tokens left in the token graph. The reason that we only keep twenty pools in the token graph is that it is hard to calculate the maximal total arbitrage profit in a graph with more pools using the Python package cvxpy\footnote{cvxpy is a Python package for solving convex optimization problems}. After token filtering, there are 11 tokens and 18 liquidity pools in total in the token graph, which is shown in Fig.\ref{token_graph}

\section{Efficiency of DEX and Token Routing Algorithm's Effects}\label{efficiency_measure_token_routing}

\subsection{Measuring the Efficiency of DEX}
Based on the data description in Section \ref{data_description}, the valid data range from June 21, 2024, to November 11, 2024. We first construct the token graph with the tokens' pool reserve data up to June 21, 2024. The constructed token graph is as shown in Fig.\ref{token_graph}. Using a similar way to store data as in \cite{zhang2025line}, the tokens' reserve information is stored in the edges.

\begin{figure}[htbp]
    \centering
    \includegraphics[width=1\linewidth]{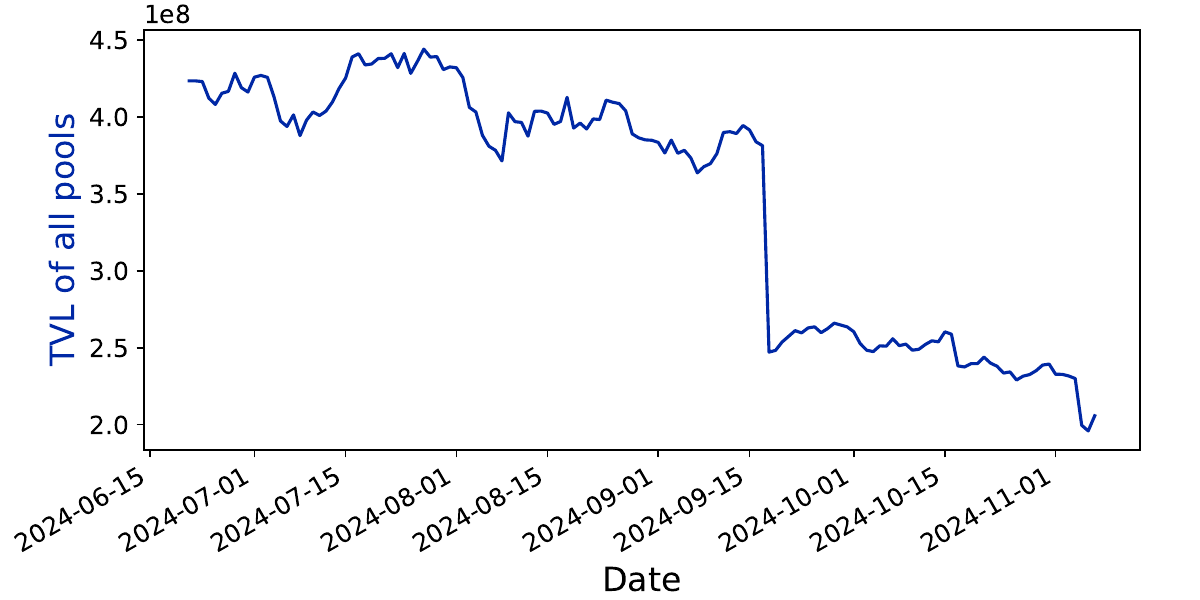}
    \caption{TVL of liquidity pools in the token graph. The TVL of a liquidity pool is calculated as the summation of the products of tokens' reserves and tokens prices.}
    \label{tvl_pool}
\end{figure}

Then, for other dates, we keep the graph connection structure the same, but only update the pools' daily reserve information stored in each edge. Firstly, we calculate the TVL of all liquidity pools in the token graph, which is shown in Fig.\ref{tvl_pool}. On September 16, 2024, there is a sharp drop in TVL coming from some liquidity providers' liquidity withdrawal.

\begin{figure}[htbp]
    \centering
    \includegraphics[width=1\linewidth]{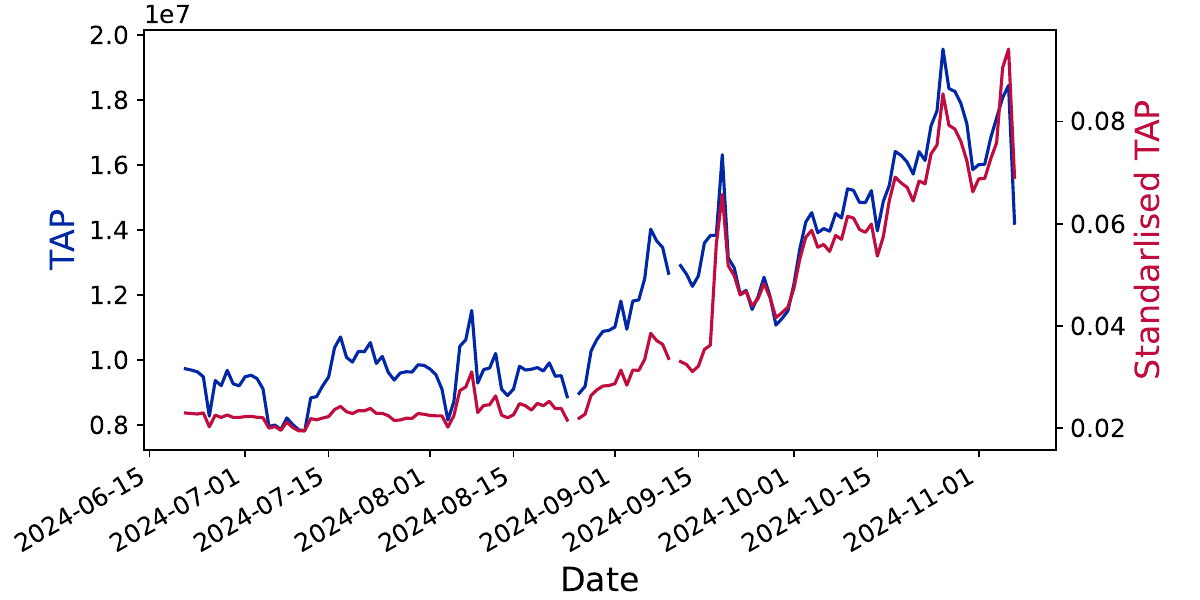}
    \caption{Total arbitrage profit (TAP) of the token graph. Standardized TAP =$ \frac{TAP}{TVL}$. }
    \label{tap}
\end{figure}

Then, we calculate the total arbitrage profit (TAP) existing in the token graph with the convex optimization method and the standardized TAP introduced in Section \ref{efficiency_measurement}. The TAP and standardized TAP from June 21, 2024, to November 8, 2024, are shown in Fig.\ref{tap}. 

From Fig.\ref{tap}, we find that both TAP and standardized TAP increased with time, which means that the Uniswap V2 is less and less efficient with time from June 21, 2024, to November 8, 2024. 


Based on the above calculation, we can find that the standardized TAP should be a good indicator of DEX efficiency. In the following section, we describe how a more profitable token routing algorithm affects the efficiency of DEX and related stakeholders' benefits.

\subsection{Token Routing Method's Effect on DEX Efficiency and Stakeholders' Benefits}

DEXs provide token routing services for traders to facilitate their trading, with token routing meaning the procedure of detecting the trading paths for token trading. For swapping a pair of tokens in DEXs, different token routing algorithms may provide different trading paths, leading to comparatively different profits for traders.
The depth-first-search (DFS) algorithm is a popular token routing method because of its low complexity, and has been used by DEXs, such as Uniswap V2. As introduced in \cite{zhang2025line}, Uniswap V2 used a linear DFS algorithm, and later they updated it to the scenario of routing splitting to improve traders' profits. 

\cite{zhang2025line} provided a more profitable token routing algorithm in DEX, called the line-graph-based method, in the case of linear routing compared to the linear DFS routing algorithm. The basic idea of this line-graph-based method is to upgrade the token graph to its line-graph to have more token combinations.

To research how different token routing algorithms affect DEX's efficiency and stakeholders' benefits, we conduct a numerical simulation experiment based on the pool reserves' statuses on June 21, 2024, which is designed as follows:
\begin{enumerate}
    \item One thousand token pairs are generated in order randomly. In each token pair, the first token is the source token to sell, and the second token is the target token to buy with the source token. The number of source token to sell is set as $\frac{M_s}{P_s}$, where $M_s$ is the monetized input and $M_s=1000\$$; $P_s$ is the source token's CEX price on June 21, 2024.

    \item For each pair of tokens, we firstly calculate the number of source tokens to input and then detect the routing path with the linear DFS algorithm from the source token to the target token. Secondly, we execute the trade and update the corresponding pools' reserves in the trading path. At the same time, we also calculate and record the TVL of the total token graph after trading, the monetized value of the target token received -the product of the target token' price $P_t$ and the number of target tokens received, and the total arbitrage profit (TAP) existing in the total token graph.

    \item For the same sequence of generated token pairs, we repeat the above procedure with the line-graph-based method.
    \label{design}
\end{enumerate}

\begin{figure}[htbp]
    \centering
    \includegraphics[width=1\linewidth]{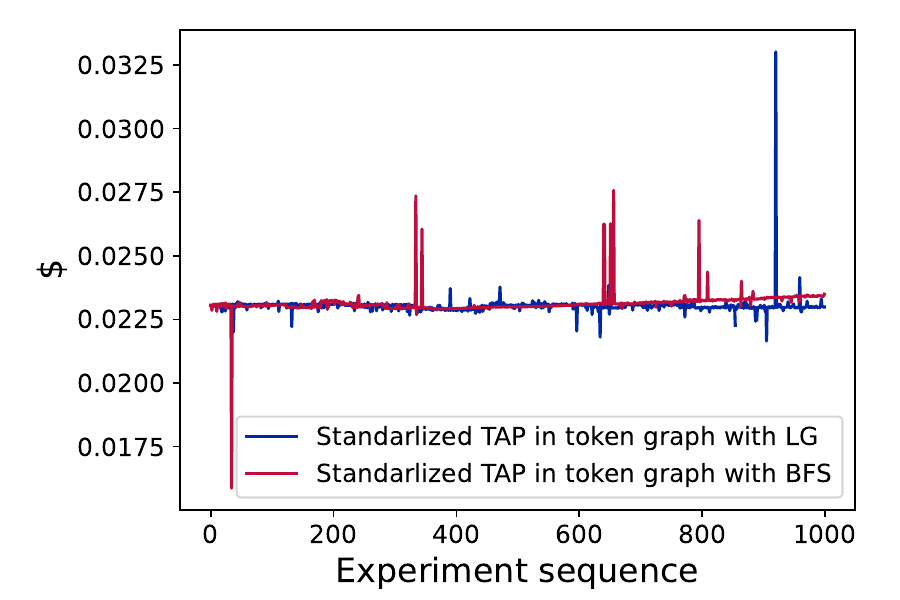}
    \caption{Standardized TAP with the linear DFS and line-graph-based routing algorithms. LG means the line-graph-based method.}
    \label{std_tap}
\end{figure}

By conducting the simulation experiment as shown above, we calculate how TAP and standardized TAP change, which is shown in Fig.\ref{std_tap} and Fig.\ref{tap}.

\begin{figure}[htbp]
    \centering
    \includegraphics[width=1\linewidth]{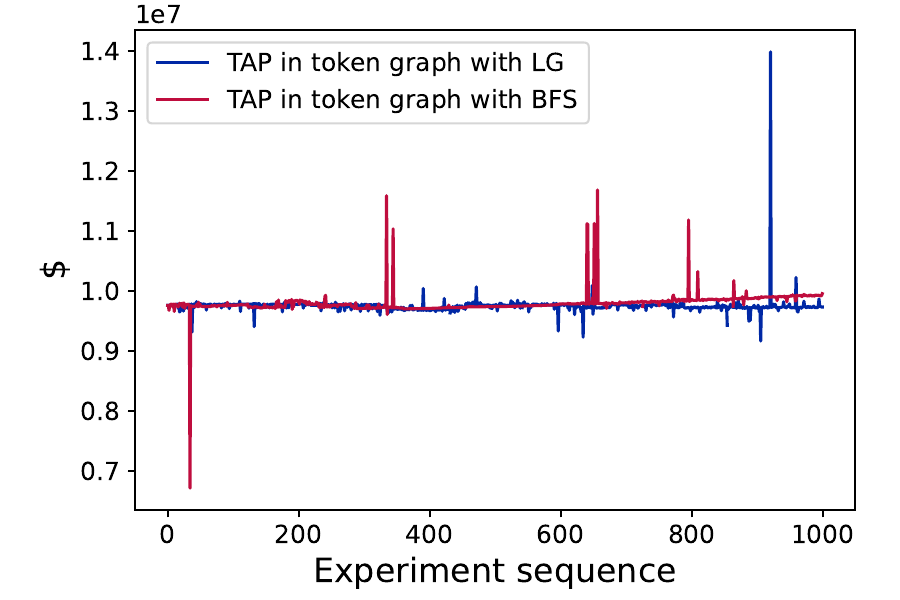}
    \caption{TAP with the linear DFS and line-graph-based routing algorithms. LG means the line-graph-based method.}
    \label{normal_tap}
\end{figure}

Fig.\ref{std_tap} and Fig.\ref{tap} show that the TAP and the standardized TAP by using the line-graph-based (LG) method are lower than those by using the DFS routing algorithm after around seven hundred token trading. A lower TAP and a standardized TAP mean a more efficient market. So, the above result indicates that a more profitable routing algorithm can help to improve the DEX's efficiency. 

\begin{figure}[htbp]
    \centering
    \includegraphics[width=1\linewidth]{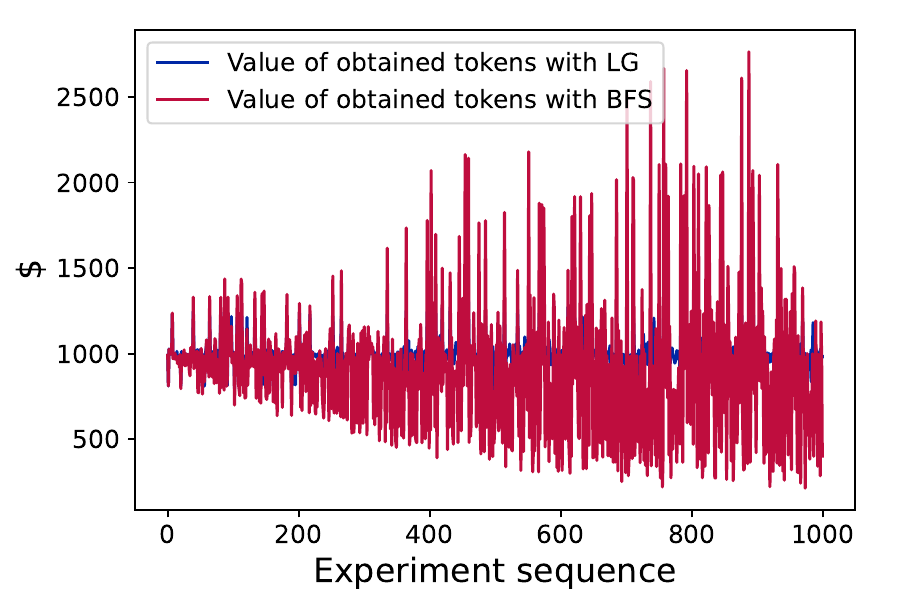}
    \caption{Monetized value of target token received in each trading. The monetized value of the source token is 1000\$. Then the number of source tokens input is $\frac{M}{P_s}$, with $M=1000$ being the amount of capital to input and $P_s$ being the source token's CEX price. }
    \label{target_token_value}
\end{figure}

We also record the monetized value of the target token received in each trading, which is shown in Fig.\ref{target_token_value}. From this figure, we find that the monetized value of the target token received by using the line-graph-based method is very stable around 1000\$. While the monetized value of the target token received by using the BFS routing algorithm fluctuates a lot, with most of them far below 1000\$ and some of them far above 1000\$. This means that the DFS algorithm may induce a loss to most traders, through some may benefit a lot because of the high price disparity they detect by chance. This also means that a DEX using a routing algorithm with poor performance in detecting profitable paths may cause more arbitrage opportunities in DEX.

\begin{figure}[htbp]
    \centering
    \includegraphics[width=1\linewidth]{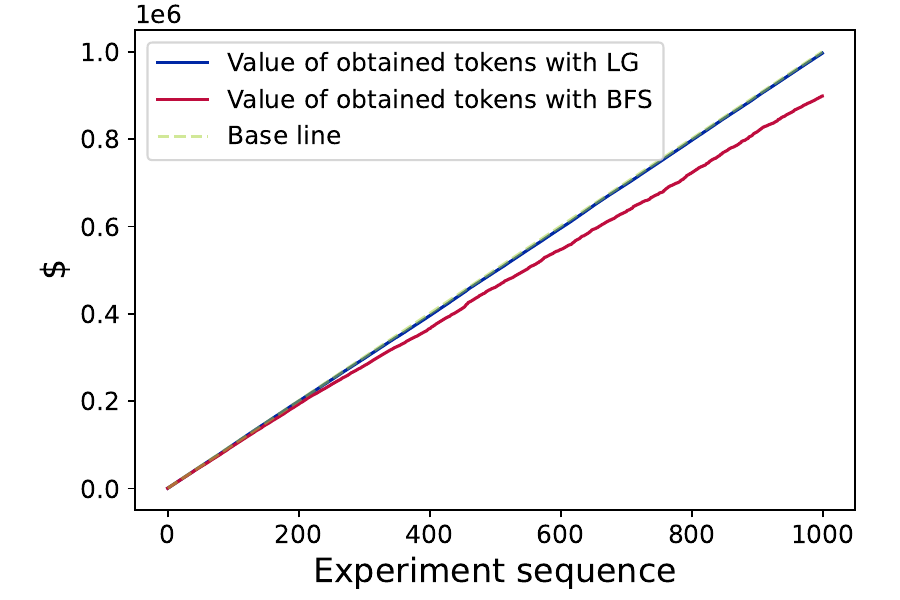}
    \caption{Accumulation of all monetized value of the target token received.}
    \label{acc_target_token_value}
\end{figure}

\begin{figure}[htbp]
    \centering
    \includegraphics[width=1\linewidth]{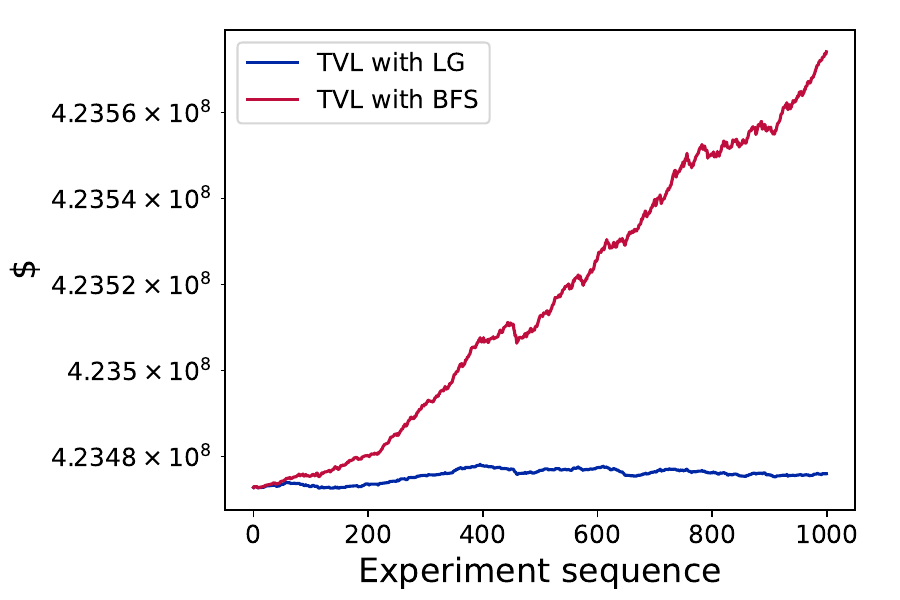}
    \caption{TVL of all liquidity pools by using the BFS and the line-graph-based Method}
    \label{total_tvl}
\end{figure}

We also calculate the accumulative monetized value of target tokens received in all previous trades and the result is shown in Fig.\ref{acc_target_token_value}.
The accumulative monetized value of target tokens received in all previous trades reflects all traders' benefit as a whole.
By Fig.\ref{acc_target_token_value}, we can find that the accumulative monetized value of target tokens received is almost equal to the amount of all capital input for all traders by using the line-graph-based method. The accumulative monetized value of target tokens received by using DFS is much lower than that using the line-graph-based method and baseline, which means that using DFS routing algorithms may lead to a loss to all traders as a whole. Then, where is the loss? We calculate the TVL of all tokens in all liquidity pools in the token graph after each trading, whose result is shown in Fig.\ref{total_tvl}.

As shown in Fig.\ref{total_tvl}, the total TVL of all liquidity pools keeps growing by using the DFS routing algorithm, which means a gain for liquidity providers as a whole but a loss for traders as a whole. While the total TVL of all liquidity pools is almost constant by using the line-graph-based method. 
The summation of TVL of all liquidity pools and the monetized value of source tokens are constant. If the monetized value of target tokens received is lower, then the TVL of all liquidity pools will be higher, which explains the results above.

\section{Summary and Discussion}

In this paper, we first put forward an indicator to measure the DEX efficiency, namely the standardized total arbitrage profit (STAP), which is the ratio between the potential total arbitrage profit (TAP) of a token graph and the TVL of all liquidity pools of the same token graph. Then, we proved that there will be no cyclic arbitrage opportunities in DEX and no arbitrage opportunities between DEX and CEX if the trade order is executed, which is calculated by solving the convex optimization problem to calculate the maximal arbitrage profits. 

Then we use the DEX efficiency indicator to measure the efficiency of a token graph with pool reserve data from Uniswap V2 and found that its efficiency decreases from June 21, 2024, to November 8, 2024. 

At last, we analyze the routing algorithm's effect on the efficiency of DEX and related stakeholders' benefits. The line-graph-based method introduced in \cite{zhang2025line} can help to detect a more profitable trading path for traders than the DFS algorithms in the linear case, which is used by Uniswap V2. If traders can obtain more target tokens by trading along one specific path over other paths, then the specific path is called a more profitable path.
Then, in the simulation, we compare the two token routing algorithms' effect on the efficiency of DEX and the related stakeholders' benefits. We find that the line-graph-based method can help to improve the efficiency of the DEX more than the DFS algorithm. The monetized value of received target tokens by traders using the line-graph-based method is almost equal to the monetized value of the source token they invest in for the trading. However, the monetized value of received target tokens by traders using the DFS algorithm fluctuates a lot. Some of them are far larger than the monetized value of the source tokens, while most of them are much lower. If we take all traders as a whole sector, we find that the accumulative monetized value of the target tokens that traders receive is almost equal to the monetized value of the source tokens that traders input if the line-graph-based routing method is applied. However, this accumulative monetized value of traders received by using the DFS algorithm is much lower comparatively. We also calculate the TVL of all liquidity pools in our token graph and find that the TVL is almost constant if we use the line-graph-based method. The TVL increases almost linearly with more token trading if the DFS algorithm is used for token routing, which means that the liquidity pool providers will benefit as a whole sector from a poor routing algorithm.
By this simulation, we can also find that the liquidity providers and traders compete to some extent, because traders' loss will be a gain for liquidity providers and vice versa.

The DEX efficiency indicator put forward in this paper can be used to measure other DEXs' efficiency and may need more tests. In this paper, we token graph only include 11 tokens and 18 liquidity pools because of the limitation of the cvxpy package. If a more accurate and powerful method can be used to solve the convex optimization problem, then our research result will be more general and robust. When we research token routing algorithms' effect on the DEX efficiency and stakeholders' benefits, we simulate a sequence of random token trading and keep the tokens' CEX prices unchanged. This is to make our analysis simpler and easier to compare. In future research, we can use real trading data for this comparison.
The research method in this paper can be extended to other DEXs, like Uniswap V3, Sushiswap, PancakeSwap, etc.. There may be other indicators that can be used to measure the efficiency DEXs, like measuring the difference of tokens prices between DEX and CEX as shown in the Appendix, and ways to research token routing algorithm's effects, which can be a research topic in the future.

\bibliographystyle{IEEEtran}
\bibliography{reference.bib}

\appendix
\subsection{Calculating Tokens' Price in DEX Using Matrix Analysis and Measuring the DEX Efficiency by the Prices' Difference against CEX}\label{appendix}

One problem with the convex optimization method is its high computational complexity and its inaccuracy when the problem scale is large.
Here, we propose the matrix analysis to measure the DEX efficiency. 
Assuming that the matrix $A$ is the exchange rate matrix with each entry equal to the exchange rate between the corresponding pair of tokens. Then we should have the following equation if the market is efficient.
\begin{equation}
    A\cdot X = \lambda \cdot X
    \label{eigen_eq}
\end{equation}
where $X$ denotes the intrinsic value of each token, $\lambda$ is the eigenvalue. We can rewrite it as the following formula.
\begin{align*}
\left[\begin{array}{cccc}
  1 & \frac{e_1}{e_2} &\cdots & \frac{e_1}{e_n} \\
 \frac{e_2}{e_1} &1 &\cdots & \frac{e_2}{e_n} \\
\frac{e_3}{e_1} &\frac{e_3}{e_2} &\cdots & \frac{e_3}{e_n} \\
  \frac{e_n}{e_1} &\frac{e_n}{e_2} &\cdots  &1 \\ 
\end{array}\right]
\begin{bmatrix}
  e_1  \\
  e_2  \\
  \vdots \\
  e_n
\end{bmatrix} 
{} &= n\cdot \begin{bmatrix}
  e_1  \\
  e_2  \\
  \vdots \\
  e_n
\end{bmatrix}
\label{matrix}
\end{align*}

where $e_1$, $e_2$,...,$e_n$ are the intrinsic value of each token and $\frac{e_i}{e_j}$ are the exchange rate between a pair of token $i$ and $j$. When the market is absolutely efficient, namely, there is no arbitrage profit, then the maximal eigenvalue equals the number of tokens $n$ and each entry of the corresponding eigenvector equals the intrinsic value of each token.

When the arbitrage exists in the market, we can still use the eigenvector to measure the intrinsic value of each token in DEXs. However, we can also get the tokens' price information from the centralized market, and these tokens' prices may be seen as the intrinsic value of those tokens in CEXs. The difference between these two price vectors can be seen as an indicator of the efficiency of DEXs.
This is just an idea and needs to be tested.
\end{document}